\documentclass[10pt,aps,prd,showpacs,eqsecnum,nofootinbib,
amsmath,amssymb,floats]{revtex4}

\DeclareMathOperator{\Order}{\mathcal{O}}

\allowdisplaybreaks

\def\nl{\\ & \quad}
\def\nlq{\\ & \qquad}

\def\picSin{((\mathbf{p}_1 \times \mathbf{S}_1) \cdot \mathbf{n}_{1 2})}
\def\piicSin{((\mathbf{p}_2 \times \mathbf{S}_1) \cdot \mathbf{n}_{1 2})}
\def\piicSiin{((\mathbf{p}_2 \times \mathbf{S}_2) \cdot \mathbf{n}_{1 2})}
\def\picSiin{((\mathbf{p}_1 \times \mathbf{S}_2) \cdot \mathbf{n}_{1 2})}
\def\SiSii{(\mathbf{S}_1 \cdot \mathbf{S}_2)}
\def\pipii{(\mathbf{p}_1 \cdot \mathbf{p}_2)}
\def\Sipi{(\mathbf{S}_1 \cdot \mathbf{p}_1)}
\def\Siipii{(\mathbf{S}_2 \cdot \mathbf{p}_2)}
\def\Sipii{(\mathbf{S}_1 \cdot \mathbf{p}_2)}
\def\Siipi{(\mathbf{S}_2 \cdot \mathbf{p}_1)}
\def\Sin{(\mathbf{S}_1 \cdot \mathbf{n}_{1 2})}
\def\Siin{(\mathbf{S}_2 \cdot \mathbf{n}_{1 2})}
\def\pin{(\mathbf{p}_1 \cdot \mathbf{n}_{1 2})}
\def\piin{(\mathbf{p}_2 \cdot \mathbf{n}_{1 2})}
\def\pipi{\mathbf{p}_1^2}
\def\piipii{\mathbf{p}_2^2}
\def\po{\mathbf{p}_1}
\def\pii{\mathbf{p}_2}

\begin{document}

\title{The next-to-leading order gravitational spin(1)-spin(2) dynamics in Hamiltonian form}

\author{Jan Steinhoff, Steven Hergt, and Gerhard Sch\"afer}
\affiliation{Theoretisch-Physikalisches Institut,
Friedrich-Schiller-Universit\"at,
Max-Wien-Pl.~1, 07743 Jena, Germany}

\date{\today}

\begin{abstract}
Based on recent developments by the authors a
next-to-leading order spin(1)-spin(2) Hamiltonian is derived
for the first time. The result is obtained within the
canonical formalism of Arnowitt, Deser, and Misner (ADM)
utilizing their generalized isotropic coordinates.
A comparison with other methods is given.

\end{abstract}

\vspace{2ex}
\noindent
\pacs{04.25.-g, 04.25.Nx}

\maketitle

\vspace{2ex}

\section{Introduction}

In this paper we present the next-to-leading (NLO) order gravitational
spin(1)-spin(2) dynamics in Hamiltonian form.
The result is based on the ADM canonical formalism 
\cite{ADM62} for spinning classical objects recently derived by Steinhoff, Sch\"afer, and Hergt
\cite{SSH07} which already has shown its power by the derivation of
the Hamiltonian of two spinning compact bodies with next-to-leading
order gravitational spin-orbit coupling, lately obtained by Damour,
Jaranowski, and Sch\"afer \cite{DJS07}.

The following notations will be used throughout the paper:
The canonical spin tensor of the $a$-th object (particle)
is the euclidean spin tensor $S_a^{i j} =S_{ai j} = \epsilon^{i j k} S_a^k $, ${\bf S}_a
= (S_a^k)=(S_{ak})$, and it also holds by definition
$\hat{S}_{ai j} = e_{ik} e_{jl} S_{ak l}$ with $e_{ik}$ being the symmetric root of
the symmetric 3-metric $\gamma_{ik}$. $\hat{S}_{ai j}$ and $S_{ai j}$
fulfill the conserved-length relations
$\hat{S}_{ai j} \gamma^{i k} \gamma^{j l} \hat{S}_{ak l}= S_{ai j}S_{ai
j}$ = const. The mass parameter of the $a$-th particle is denoted
$m_a$.  The 4-vector $n^{\mu}$ is the unit vector orthogonal to
the spacelike hypersurfaces $t$ = const.; its components are
$n_{\mu} = (-N, 0,0,0)$, where $N$ is the lapse function. The short-cut
notation $-np_a$ is used for
$\sqrt{m_a^2 + \gamma^{ij}p_{ai}p_{aj}}$, where $(p_{ai}) = {\bf p}_a$ denotes the
canonical momentum of the $a$-th particle. The canonical particle position
variables are $(x^i_a) = {\bf x}_a$ and the velocities read ${\bf v}_a= (v^i_a) =
(\dot{x}^i_a)$, where the dot means coordinate time derivative.

Our units are $c=1$, where $c$ is the velocity of light.
$G$ will denote the Newtonian gravitational constant. Greek indices will run over $0,1,2,3$, Latin over
$1,2$ if from the beginning of the alphabet and $1,2,3$ if from the middle.
For the signature of spacetime we choose +2.

\section{Spinning objects in the ADM formalism}

Recently in \cite{SSH07} it has been shown that the matter source parts
of the energy and momentum constraint equations, respectively
$\mathcal{H}^{\text{M}}$ and $\mathcal{H}_i^{\text{M}}$,
are given in terms of canonical position, momentum, and spin variables
in the form, to the post-Newtonian orders indicated,
\begin{align}\label{Hmatter}
	\mathcal{H}^{\text{M}} &= \sum_a \left[ - np_a \delta_a\left\{1
	+ \frac{1}{2}\frac{p_{a j} \hat{S}_{a li} \gamma^{lk} \gamma^{i j}_{~ , k} }{(np_a)^2 }\right\}
        + \left( \frac{p_{a l} \hat{S}_{aij} \gamma^{il}  \gamma^{kj} \delta_a}{ m_a - np_a } \right)_{, k} \right]+ \Order{(S/c^6)}\,, \\
	\mathcal{H}^{\text{M}}_i &= \sum_a \left[ p_{a i} \delta_a +
        \frac{1}{2}\left({S}_{a i k}
		+  \frac{p_{a l} {S}_{a l ( i} p_{a k )}}{ m_a^2 }\right)\delta_{a, k}  \right]+ \Order{(S/c^4)}\,.
\end{align}
These expressions are sufficient for the
derivation of the Hamiltonian in \cite{DJS07}. The applied equal-time Poisson brackets read
\begin{equation}\label{minkowski_canonical_algebra}
	\{ x^i_a , p_{aj} \} = \delta_{ij}\,, \qquad 
	\{ {S}^{i j}_a , {S}^{k l}_a \} = \delta_{i k} {S}^{j l}_a - \delta_{j k} {S}^{i l}_a
		- \delta_{i l} {S}^{j k}_a + \delta_{j l} {S}^{i k}_a\,, \qquad \text{zero otherwise}\,.
\end{equation}
It is important to note that the expressions for
$\mathcal{H}^{\text{M}}$ and $\mathcal{H}_i^{\text{M}}$
are given in the ADM transverse-traceless (ADMTT) gauge which
refers to generalized isotropic coordinates defined by the conditions
\begin{equation}{\label{ADM}}
\gamma_{ij}=\left(1+\frac{1}{8}\phi\right)^{4}\delta_{ij}+h_{ij}^{\text{TT}}\,, \quad \pi^{ii}=0\,,
\end{equation}
with $h_{ij}^{\text{TT}}$ being transverse and traceless ($h_{ij,j}^{\text{TT}}=0$, $h_{ii}^{\text{TT}}=0$).
The square root of the metric takes the form, to sufficient approximation for the Hamiltonian,
\begin{equation}
e_{ij}  = \left(1 + \frac{1}{4}\phi\right)\delta_{ij} +
\Order\left(\phi^2, h_{ij}^{\text{TT}} \right)\,.
\end{equation}
In this approximation, $\hat{S}^{~~j}_{a i} \equiv \hat{S}_{a
il}\gamma^{lj} = \hat{S}_{alj}\gamma^{li} \equiv \hat{S}^i_{aj} =S_{a ij}=S_a^{ij}$ holds.

The canonical conjugate to $h_{ij}^{\text{TT}}$,  $\frac{1}{16\pi
G}\pi^{ij}_{\text{TT}}$, i.e.,
\begin{equation}
\frac{1}{16\pi G} \{h_{ij}^{\text{TT}}({\bf x},t), \pi^{kl}_{\text{TT}}({\bf x}',t)\}
= \delta^{\text{TT}kl}_{ij}({\bf x} - {\bf x}')\,, \qquad \text{zero otherwise}\,,
\end{equation}
will play no role in the calculations of the present paper. Only the longitudinal
part of  $\pi^{ij}$, $\tilde{\pi}^{ij}$, contributes (notice: $\pi^{ij} = \tilde{\pi}^{ij}  + \pi^{ij}_{\text{TT}}$).
Thus, the ADM Hamiltonian $H$ will not depend on
$\pi^{ij}_{\text{TT}}$, i.e.,
\begin{equation}
	H[{\bf x}_a, {\bf p}_a, h_{ij}^{\text{TT}}] = - \frac{1}{16\pi G}\int{ \text{d}^3 x \, \Delta \phi}\,.
\end{equation}
Therefore, $h_{ij}^{\text{TT}}$ is allowed to be
replaced by matter variables (after solution of the evolution equations
which in our approximation is an elliptic equation only)
to get an autonomous Hamiltonian. Otherwise the transition to a Routhian would have to be
performed, see \cite{JS97}.
The equations to be solved for the obtention of the autonomous matter
Hamiltonian read, on the one side
\begin{gather}
\frac{1}{\sqrt{\gamma}}\left(-\gamma_{ik}\gamma_{jl}\pi^{ij}\pi^{kl}+\frac{1}{2}\gamma_{ij}\gamma_{kl}\pi^{ij}\pi^{kl}\right)
+ \sqrt{\gamma}R_{(3)}=16\pi G\mathcal{H}^{\text{M}} \,, \\
-\gamma_{i j} \pi^{jk}_{~~;k} =8\pi G \mathcal{H}^{\text{M}}_i\,,
\end{gather}
where $R_{(3)}$ denotes the Ricci scalar of the $t$ = const.\ slices,
$\gamma$ is the determinant of the 3-metric, and ; the
3-dim.\ covariant derivative, and on the other side
\begin{equation}
0=\{H, \pi^{ij}_{\text{TT}}\} + \Order{(1/c^6)}\,.
\end{equation}
After tedious calculations, the NLO spin(1)-spin(2)
interaction part of the Hamiltonian results in
\begin{equation}\label{H2PN_SS}
\begin{split}
	H_{\text{SS}}^{\text{NLO}} &=
		\frac{1}{2 m_1 m_2 r_{1 2}^3} [
			\tfrac{3}{2} \picSin \piicSiin
			+ 6 \piicSin \picSiin \nlq
			- 15 \Sin \Siin \pin \piin
			- 3 \Sin \Siin \pipii \nlq + 3 \Sipii \Siin \pin
			+ 3 \Siipi \Sin \piin + 3 \Sipi \Siin \piin \nlq
			+ 3 \Siipii \Sin \pin - \tfrac{1}{2} \Sipii \Siipi
			+ \Sipi \Siipii \nlq - 3 \SiSii \pin \piin + \tfrac{1}{2} \SiSii \pipii
		] \nl
		+ \frac{3}{2 m_1^2 r_{1 2}^3} [
			- \picSin \picSiin
			+ \SiSii \pin^2 - \Sin \Siipi \pin
		] \nl
		+ \frac{3}{2 m_2^2 r_{1 2}^3} [
			- \piicSiin \piicSin
			+ \SiSii \piin^2 - \Siin \Sipii \piin
		] \nl
		+ \frac{6 ( m_1 + m_2 )}{r_{1 2}^4} [ \SiSii - 2 \Sin \Siin ] \,,
\end{split}
\end{equation}
where $r_{12}= |{\bf x}_1-{\bf x}_2|$ is the euclidean distance between the two
particles and ${\bf n}_{12}$ denotes the unit vector $r_{12} {\bf n}_{12}= {\bf x}_1 - {\bf x}_2$.

\section{Different derivation of the spin(1)-spin(2) Hamiltonian}

We follow here the procedure described in \cite{DJS07}.
The implementation of spin into the Eq.\ (4.9) of \cite{DJS07} results in
\begin{equation}
	v^{i \, \text{spin}}_{(3) a} =
		- \sum_{b \neq a} \left( \frac{3 m_b S_{a ij}}{2 m_a} + 2 S_{b ij} \right) \frac{n_{a b}^j}{r_{a b}^2}\,.
\end{equation}
The NLO spin(1)-spin(2) interation Hamiltonian is given by
\begin{equation}
	H_{\text{SS}}^{\text{NLO}} = \tilde{\Omega}_{(4){ij}}\,S_{1}^{i}\,S_{2}^{j}
		= \mathbf{\Omega}_{(4)}^{\text{spin(2)}} \cdot \mathbf{S}_1
		= \mathbf{\Omega}_{(4)}^{\text{spin(1)}} \cdot \mathbf{S}_2 \,.
\end{equation}
Using Eq.\ (4.10b) in Ref.\ \cite{DJS07}, but calculated for metric functions resulting from our
matter source terms, the obtained Hamiltonian coincides
with the one calculated in the present paper.

In this completely independent approach also the matter source part
$\frac{1}{2} N \gamma^{i k} \gamma^{j l} T_{kl}$ of the evolution equations contributes,
\begin{equation}
	T_{kl} = \sum_a \left[-\frac{p_{a k} p_{a l}}{np_a} \delta_a
		- \frac{{S}_{aj ( k} p_{a l )}}{m_a} \delta_{a , j} \right]
		+ \Order{(S/c^2)}\,,
\end{equation}
and lapse and shift functions have to be determined too. Details of the
calculations can be found in \cite{SSH07}.

\section{Consistency}

The correctness of the derived spin(1)-spin(2) Hamiltonian can best be
verified by the construction of  the global Poincar\'e algebra.
The generators of  the global Poincar\'e algebra are the total linear
momentum ${\bf P} = {\bf p}_1+ {\bf p}_2$, the total angular momentum
${\bf J} = {\bf x}_1 \times {\bf p}_1+ {\bf x}_2 \times {\bf p}_2 + {\bf
S}_1 +{\bf S}_2$, the total Hamiltonian $H$, and the total center-of-mass generator
${\bf G}$. The latter object is defined by $\mathbf{G} = - \frac{1}{16\pi
G}\int{ \text{d}^3 x \, \mathbf{x} \Delta \phi}$, e.g., see \cite{RT74},
and turns out to be
\begin{equation}
 \mathbf{G} = \mathbf{G}_{\text{PM}} + \mathbf{G}_{\text{SO}}+\frac{G}{2
 r_{12}^2}\left[\Siin\mathbf{S}_{1}-\Sin\mathbf{S}_{2} + \left(3\Siin\Sin-\SiSii\right)\frac{\mathbf{x}_{1}+\mathbf{x}_{2}}{r_{12}}\right]\,,
\end{equation}
where $\mathbf{G}_{\text{SO}}$ denotes the NLO spin-orbit coupling contribution
as given in \cite{DJS07} and $\mathbf{G}_{\text{PM}}$ is the point-mass part from
\cite{DJS00}. It is straightforward to show that the above
generators do fulfill the Poincar\'e algebra.

\section{Comparison with other methods and results}

In a recent paper \cite{PR06}, based on Ref.\ \cite{P06}, Porto and
Rothstein derived a next-to-leading order spin(1)-spin(2) potential
using an action approach with spin supplementary conditions imposed
on the action level. A consistency calculation, however, which would have
shown that the spin supplementary conditions are preserved under the variational principle, and thus
under the equations of motion, has not been undertaken nor has it been shown
that, in contrast to their claim, the used position, velocity, and spin
variables are those that relate to canonical ones in standard manner,
cf., e.g., \cite{YB93}. A consistency check of their intuitive canonical approach
by an independent method is therefore necessary, which is the subject of
this Section.

The relevant part of the Lagrangian of Porto and Rothstein reads $L^{\text{PR}} =
\frac{1}{2} m_1 \mathbf{v}_1^2 + \frac{1}{2} m_2 \mathbf{v}_2^2 - V_N -
V_{\text{SO}}^{\text{LO}} - V_{\text{SS}}^{\text{PR}}$ where
$V_N$ is the Newtonian potential, $V_{\text{SS}}^{\text{PR}}$ is
given by Eq.\ (12) in \cite{PR06} and the leading order spin-orbit
coupling potential function $V_{\text{SO}}^{\text{LO}}$ results
from, e.g., \cite{DJS07},
\begin{equation}
	V_{\text{SO}}^{\text{LO}} = \frac{G}{r_{1 2}^2} \bigg[
		\frac{3}{2} m_2 ((\mathbf{v}_1 \times \mathbf{S}_1) \cdot \mathbf{n}_{1 2})
		- 2 m_2 ((\mathbf{v}_2 \times \mathbf{S}_1) \cdot \mathbf{n}_{1 2}) %\nl
		- \frac{3}{2} m_1 ((\mathbf{v}_2 \times \mathbf{S}_2) \cdot \mathbf{n}_{1 2})
		+ 2 m_1 ((\mathbf{v}_1 \times \mathbf{S}_2) \cdot \mathbf{n}_{1 2})
		\bigg]\,.
\end{equation}
The canonical momenta are given by $\mathbf{p}_a = \frac{\partial
L^{\text{PR}}}{\partial \mathbf{v}_a}$, i.e.,
\begin{align}
	\po &= m_1 \mathbf{v}_1 - \frac{G}{r_{1 2}^2} \bigg[
			\frac{3}{2} m_2 (\mathbf{S}_1 \times \mathbf{n}_{1 2})
			+ 2 m_1 (\mathbf{S}_2 \times \mathbf{n}_{1 2}) \bigg]
		- \frac{\partial V_{\text{SS}}^{\text{PR}}}{\partial \mathbf{v}_1}, \\
	\pii &= m_2 \mathbf{v}_2 + \frac{G}{r_{1 2}^2} \bigg[
			\frac{3}{2} m_1 (\mathbf{S}_2 \times \mathbf{n}_{1 2})
			+ 2 m_2 (\mathbf{S}_1 \times \mathbf{n}_{1 2}) \bigg]
		- \frac{\partial V_{\text{SS}}^{\text{PR}}}{\partial \mathbf{v}_2}\,.
\end{align}
The Hamiltonian of Porto and Rothstein then takes the form $H^{\text{PR}} = H_N +
H_{\text{SO}}^{\text{LO}} +
H_{\text{SS}}^{\text{PR}}$. We can get
$H^{\text{PR}}$ by replacing the velocities by canonical
momenta in the following expression
\begin{equation}
	H^{\text{PR}} = \mathbf{v}_1 \cdot \po + \mathbf{v}_2 \cdot \pii - L^{\text{PR}}\,.
\end{equation}
Note that the $\frac{\partial V_{\text{SS}}^{\text{PR}}}{\partial \mathbf{v}_a}$ terms do not contribute.
The difference between $H_{\text{SS}}^{\text{NLO}}$ and $H^{\text{PR}}_{\text{SS}}$ reads
\begin{equation}
\begin{split}
	\delta H_{\text{SS}}^{\text{NLO}} &= \frac{G}{2 m_1 m_2 r_{1 2}^3} [
			3 \Sipii \Siin \pin + 3 \Siipi \Sin \piin \nlq
			- 2 \Sipii \Siipi - 6 \SiSii \pin \piin + 2 \SiSii \pipii
		] \nl
		+ \frac{G}{2 m_1^2 r_{1 2}^3} [
			3 \SiSii \pin^2 - \SiSii \pipi
			- 3 \Sin \Siipi \pin + \Sipi \Siipi
		] \nl
		+ \frac{G}{2 m_2^2 r_{1 2}^3} [
			3 \SiSii \piin^2 - \SiSii \piipii
			- 3 \Siin \Sipii \piin + \Sipii \Siipii]\,.
\end{split}
\end{equation}
Obviously, there is agreement at $\Order{(G^2)}$.

There should exist an infinitesimal generator $g$ for a canonical
transformation such that $\delta H_{\text{SS}}^{\text{NLO}} = \{ H_N , g \}$. Plugging in the ansatz
\begin{equation}\label{ct1}
\begin{split}
	g  &= a \frac{G}{r_{1 2}^2} \left[ \frac{1}{m_1} \Sipi \Siin - \frac{1}{m_2} \Siipii \Sin \right] \nl
		+ b \frac{G}{r_{1 2}^2} \left[ \frac{1}{m_2} \Sipii \Siin - \frac{1}{m_1} \Siipi \Sin \right] \nl
		+ c \frac{G}{r_{1 2}^2} \left[ \frac{1}{m_1} \SiSii \pin - \frac{1}{m_2} \SiSii \piin \right] \nl
		+ d \frac{G}{r_{1 2}^2} \left[ \frac{1}{m_1} \Sin \Siin \pin - \frac{1}{m_2} \Sin \Siin \piin \right]
\end{split}
\end{equation}
and comparing $\Order{(G)}$ terms gives
\begin{equation}\label{ct2}
	a = 0\,, \quad
	b = \frac{1}{2}\,, \quad
	c = \frac{1}{2}\,, \quad
	d = 0\,.
\end{equation}
The vanishing of $\Order{(G^2)}$ terms yields
\begin{equation}
	c = 0\,, \quad
	- a + b - d = 0\,,
\end{equation}
which is incompatible with the canonical transformation that is needed for the $\Order{(G)}$ terms.

In Ref.\ \cite{PR07}, which is a short reply to the first version of the present paper \cite{SHS07},
Porto and Rothstein pointed out that their result in \cite{PR06} is incomplete
in the sense that it only includes contributions from spin-spin \emph{diagrams}. It was not realized
in \cite{PR06} that spin-orbit diagrams also contribute to the next-to-leading
order spin(1)-spin(2) interaction. If these contributions
are included, the canonical transformation defined by (\ref{ct1}) and
(\ref{ct2}) leads to an additional agreement at $\Order{(G^2)}$, and thus to full agreement.
It should be recalled that our derivation is completely different and
includes all contributions to the spin(1)-spin(2) interaction from the very beginning.

\acknowledgments 
This work is supported by the Deutsche Forschungsgemeinschaft (DFG) through
SFB/TR7 ``Gravitational Wave Astronomy''.

\end{document}